\documentclass[11pt,a4paper]{article}

\usepackage{graphicx}
\usepackage{caption}
\usepackage{authblk}
\usepackage[left=2.5cm,right=2.5cm,top=1.5cm,bottom=1.5cm,includeheadfoot]{geometry}

\title{Improved stabilization scheme for extreme ultraviolet quantum interference experiments}

\author[1]{Daniel Uhl}
\author[1]{Andreas Wituschek}
\author[1]{Ulrich Bangert}
\author[1]{Marcel Binz}
\author[2]{Carlo Callegari}
\author[2]{Michele Di Fraia}
\author[2]{Oksana Plekan}
\author[2]{Kevin Charles Prince}
\author[3]{Giulio Cerullo}
\author[2,4]{Luca Giannessi}
\author[2]{Miltcho Danailov}
\author[1]{Giuseppe Sansone}
\author[5,6]{Tim Laarmann}
\author[1]{Rupert Michiels}
\author[7]{Marcel Mudrich}
\author[8]{Paolo Piseri}
\author[9]{Richard James Squibb}
\author[9]{Raimund Feifel}
\author[10]{Stefano Stranges}
\author[1]{Frank Stienkemeier}
\author[1]{Lukas Bruder\thanks{lukas.bruder@physik.uni-freiburg.de}}

\affil[1]{Institute of Physics, University of Freiburg, Hermann-Herder-Str. 3, 79104 Freiburg, Germany.}
\affil[2]{Elettra-Sincrotrone Trieste S.C.p.A., 34149 Basovizza, Trieste, Italy.}
\affil[3]{IFN-CNR and Dipartimento di Fisica, Politecnico di Milano, Piazza L. da Vinci 32, 20133 Milano, Italy.}
\affil[4]{INFN Laboratori Nazionali di Frascati, Via E. Fermi 40, 00044, Frascati, Roma, Italy}
\affil[5]{Deutsches Elektronen-Synchrotron DESY, Notkestraße 85, 22607 Hamburg, Germany.}
\affil[6]{The Hamburg Centre for Ultrafast Imaging CUI, Luruper Chaussee 149, 22761 Hamburg, Germany.}
\affil[7]{Department of Physics and Astronomy, Aarhus University, Ny Munkegade 120, 8000 Aarhus C, Denmark.}
\affil[8]{Dipartimento di Fisica "Aldo Pontremoli", Università degli Studi di Milano, Via Celoria 16, 20133 Milano, Italy}
\affil[9]{Department of Physics, University of Gothenburg, Origovägen 6 B, 412 96 Gothenburg, Sweden.}
\affil[10]{University of Rome “La Sapienza”, Piazzale Aldo Moro 5, 00185 Roma, Italy.}

\date{}

\begin{document}
\maketitle

\begin{abstract}
Interferometric pump-probe experiments in the extreme ultraviolet (XUV) domain are experimentally very challenging due to the high phase stability required between the XUV pulses. 
Recently, an efficient phase stabilization scheme was introduced for seeded XUV free electron lasers (FELs) combining shot-to-shot phase modulation with lock-in detection [A. Wituschek et al., Nat Commun 11, 1 (2020)]. 
This method stabilized the seed laser beampath on the fundamental ultraviolet wavelength to a high degree. 
Here, we extend this scheme including the stabilization of the XUV beampath, incorporating phase fluctuations from the FEL high gain harmonic generation process. 
Our analysis reveals a clear signal improvement with the new method compared to the previous stabilization scheme. 
\end{abstract}

\maketitle

\section{Introduction}
\label{sec:introduction}
Interference phenomena can be exploited to control quantum pathways with high precision\,\cite{jones_multiphoton_1995} and to improve the attainable resolution of experiments with respect to temporal, spectral and spatial information\,\cite{paul_observation_2001, itatani_attosecond_2002, dantus_femtosecond_1990, gruner_vibrational_2011, dai_plasmonic_2020}. 
In order to observe interference between two optical pulses, sub-cycle phase/delay stability between the pulse replicas is required. 
This explains why techniques exploiting the interference between multiple laser pulses/beams are very challenging to perform at short wavelengths in the extreme ultraviolet (XUV) to X-ray regime where only a few such experiments have been reported to date\,\cite{prince_coherent_2016, jansen_spatially_2016, wituschek_tracking_2020, wituschek_phase_2020, kaneyasu_electron_2021, prince_coherent_2021,  usenko_attosecond_2017, usenko_auger_2020, allaria_fermi_2015, skruszewicz_table-top_2021, bellini_temporal_1998, salieres_frequency-domain_1999, cavalieri_ramsey-type_2002}. 

At visible wavelengths several phase stabilization methods have been developed to solve the phase jitter issue\,\cite{scherer_fluorescence-detected_1991, tian_femtosecond_2003, brixner_phase-stabilized_2004, volkov_active_2005, tekavec_wave_2006}. 
However, implementing these stabilization concepts in the XUV and X-ray domain is very challenging, due to the lack of suitable optics for short wavelengths and/or the required complex optical setups. 
Among the established stabilization schemes, one very efficient concept is the phase modulation (PM) technique\,\cite{tekavec_wave_2006}. 
This method features extraordinary sensitivity to probe highly dilute samples in the condensed and gas phase\,\cite{bruder_coherent_2018, bruder_coherent_2019, tamimi_fluorescence-detected_2020}, which can be further improved using selective detection schemes\,\cite{bruder_phase-modulated_2015,bruder_efficient_2015, bruder_delocalized_2019, uhl_coherent_2021}, and can be transferred to shorter wavelengths using frequency up-conversion\,\cite{bruder_phase-modulated_2017}. 
Recently, this method was successfully implemented in the XUV domain using high-gain harmonic generation (HGHG)\,\cite{yu_generation_1991} at the seeded XUV free electron laser (FEL) FERMI\,\cite{allaria_fermi_2015} where it enabled the measurement of the dephasing of a Fano resonance in the time domain\,\cite{wituschek_tracking_2020}. 
It was also implemented in tabletop high harmonic generation (HHG), which enabled the spectral characterization of narrow-band harmonics with very high spectral resolution ($< 0.7$\,meV, $\Delta E/ E \sim 10^{-5}$)\,\cite{wituschek_phase_2020}. 

In these experiments, the timing and phase control, including the phase stabilization, was implemented on the fundamental wavelength before the harmonic generation was performed. 
This is advantageous, since no modification of the XUV beampath is necessary, which greatly simplifies the experimental implementation. 
However, this approach does not stabilize phase fluctuations introduced in the harmonic generation process itself or the phase jitter introduced in the XUV beampath. 
The same applies to other demonstrated concepts stabilizing the pulse sequences on the fundamental wavelength\,\cite{bellini_temporal_1998, jansen_spatially_2016, koll_experimental_2021-1}. 
In the tabletop-approach based on HHG in gases, this problem is usually not significant, since the phase jitter introduced between the pump and probe pulses can be kept small\,\cite{salieres_frequency-domain_1999}. 
In contrast, in HGHG the phase jitter can be substantial\,\cite{gauthier_spectrotemporal_2015, wituschek_tracking_2020} and, hence, including the XUV beampath in the stabilization method would be desirable. 
The PM technique is a passive stabilization scheme, which tracks the phase fluctuations and corrects them in the signal detection, in contrast to most other stabilization methods. 
This opens up the possibility to track and correct the phase fluctuations of the harmonic generation and the XUV beampath, even if only the beampath of the fundamental wavelength is controlled. 

Here, we extend the PM technique in order to track and correct phase fluctuations in the XUV beampath at the FERMI FEL-1 source. 
This is not possible with other concepts actively stabilizing the interferometer at the fundamental wavelength. 
Our approach clearly improves the signal quality compared to the same measurement stabilizing only the beampath before the harmonic generation. 
Further improvement of the signal quality is limited by the shot-to-shot amplitude and phase fluctuations deriving from the FEL process, producing high-frequency noise in the measurement, and thus requiring self-referencing methods in order to be fully corrected. 

\section{Method}
\label{sec:method}
\begin{figure}[ht!]
\centering
\includegraphics[width=0.8\linewidth]{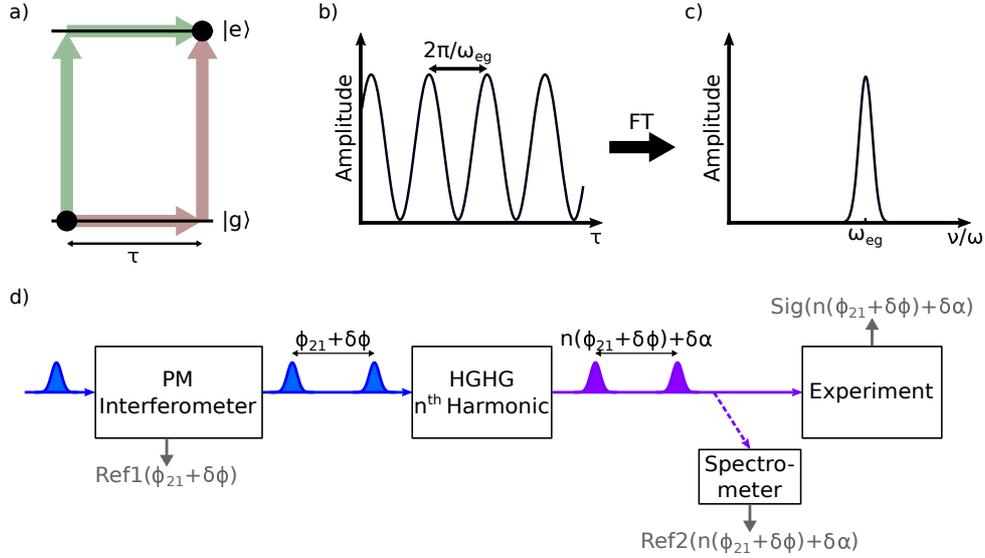}
\captionsetup{format=hang}
\caption{
Experimental scheme. 
(a) Principle of electronic WPI. 
Pump and probe pulses each excite a quantum pathway (green, brown) in the system. 
The interference of the excitation pathways gives rise to interference fringes as a function of the relative phase between the pathways. 
(b) Excited state population as a function of the pump-probe delay, reflecting the constructive/destructive pathway interference. 
(c) Fourier transform of (b) reveals the linear absorption spectrum of the system. 
(d) Experimental setup: The UV seed pulses are split-and-delayed and phase-modulated in the PM interferometer setup. 
Subsequent HGHG generates phase-modulated XUV pump-probe pulse pairs used to excite the sample. 
A portion of the XUV light is branched off and analyzed in the XUV beamline spectrometer. 
The reference for lock-in demodulation is retrieved either from the seed laser interferometer (Ref1) or from the XUV spectrometer (Ref2). 
$\phi_{21}$: phase difference between seed laser pump and probe pulses. 
$\delta \phi$: fluctuations of $\phi_{21}$. 
$\delta \alpha$: accumulated phase fluctuations from HGHG and subsequent XUV beampath. 
}
\label{fig1}
\end{figure}
The interference experiments are based on XUV electronic wave packet interferometry (WPI) (Fig.\,\ref{fig1}a--c). 
Here, pump and probe pulses each excite a coherent superposition of electronic ground and excited states in the system, giving rise to constructive/destructive interference in the excited state population as a function of the relative phase between the pulses/excitation pathways (Fig.\,\ref{fig1}a). 
Detecting the excited state population, e.g. by fluorescence or photoionization, while sweeping the temporal pump-probe delay results in an interferogram in the time domain (Fig.\,\ref{fig1}b). 
This signal reflects the free polarization decay of the induced dipole transition between ground and excited electronic states. 
Accordingly, the oscillation frequencies ($\omega_{eg}$) correspond to the energy difference between the involved states ($\omega_{eg}=(E_e-E_g)/\hbar$), and a Fourier transform yields the linear absorption spectrum of the sample (Fig.\,\ref{fig1}c). 
The attainable frequency resolution is given by the scan range in the time domain and is therefore decoupled from the spectral width of the XUV pulses, thus facilitating high resolution spectra even with ultrashort and broadband pump and probe pulses. 

In order to detect clean interferograms, the phase fluctuations between pump and probe pulses must be kept small: $\delta \phi < 2\pi/50$\,\cite{hamm_concepts_2011}, which is increasingly difficult to achieve at short wavelengths. 
In the PM technique, this issue is solved by removing the phase jitter from the signal by heterodyne detection with a reference signal exhibiting the same phase fluctuations. 
To this end, the relative carrier-envelope phase (CEP) between pump and probe pulses is modulated at a frequency of $\Omega$ on a shot-to-shot basis. 
This characteristic modulation enables lock-in amplification to efficiently extract the interference signal from background contributions. 
At the same time, the optical interference between pump and probe pulses is detected to track the phase changes and fluctuations in the interferometer. 
Using this signal as reference for the lock-in amplification returns as output the phase difference between the input signal and the reference: $\phi_\mathrm{out} = \phi_\mathrm{sig}-\phi_\mathrm{ref}$. 
This leads to cancellation of the correlated phase jitter of the signals and down-shifts the signal frequencies by orders of magnitude to a lower frequency regime (\textit{rotating frame sampling})\,\cite{tekavec_wave_2006}. 
Accordingly, the residual signal phase at the lock-in amplifier (LIA) output is
\begin{equation}\label{eq:1}
   \phi_\mathrm{out} = \Delta \omega \tau + \delta \phi_\mathrm{sig-ref}\, ,
\end{equation}
where $\Delta \omega=\omega_{eg}-\omega_\mathrm{ref}$ denotes the frequency difference between the optical resonance $\omega_{eg}$ and the optical frequency of the reference signal $\omega_\mathrm{ref}$, $\tau$ denotes the pump-probe delay and $\delta \phi_\mathrm{sig-ref}$ the uncorrelated phase noise between signal and reference. 

A detailed description of the PM technique and of the experimental setup are given in Refs.\,\cite{tekavec_wave_2006, bruder_phase-modulated_2015, wituschek_tracking_2020, wituschek_high-gain_2020, wituschek_stable_2019}. 
Fig.\,\ref{fig1}d gives an overview over the experimental setup. 
With a highly stable, monolithic PM interferometer setup\,\cite{wituschek_stable_2019} in the seed laser beampath, phase-modulated pump and probe pulses are generated. 
Upon generation of the $n$'th harmonic of the seed laser pulses, the phase of each pulse increases by an integer factor of $n$ (Fig.\,\ref{fig1}d). 
The XUV pulses excite and ionize the sample and the produced ions are mass-resolved with an ion time-of-flight (TOF) detector. 
The phase modulation introduced on the seed laser is reflected in the ion yield and is demodulated with a LIA. 
In addition, phase fluctuations $\delta \alpha$ introduced in the HGHG process and the XUV beamline contribute to the signal. 
Here, the timing jitter between the seed laser pulses and the electron bunch is identified as the major contribution to $\delta \alpha$\,\cite{wituschek_tracking_2020}. 
In the previous works\,\cite{wituschek_tracking_2020, wituschek_high-gain_2020}, the signal (Sig) was demodulated with Ref1, tracking the phase changes and fluctuations in the seed laser setup. 
To this end, Ref1 was digitized and its $n$'th harmonic was computed inside the LIA. 
In that implementation, the XUV phase fluctuations $\delta \alpha$ were not tracked and therefore did not cancel upon lock-in amplification. 
In the current work, we extend the scheme. 
Instead of using a commercial LIA for the signal detection, we employ a software-based universal lock-in amplifier (ULIA)\,\cite{uhl_flexible_2021}, which provides more flexibility in the signal processing of the experimental data and the reference signal. 

While commercial LIAs require smooth, single-channel analog signals as inputs, the ULIA can process pulsed (single-shot-resolved) multidimensional detector signals\,\cite{uhl_coherent_2021}. 
For the XUV experiments, this has two advantages. 
First, entire photoelectron/-ion TOF traces can be processed on the single-shot level for the ULIA signal input, and, thus, interference signals from many different ion/electron contributions can be extracted, simultaneously in a single measurement. 
To this end, the TOF traces are digitized with a fast analog-to-digital converter and each time bin is demodulated with the ULIA algorithm, similar to the procedure in Ref.\,\cite{uhl_coherent_2021}. 

\begin{figure}[ht!]
\centering
\includegraphics[width=0.8\linewidth]{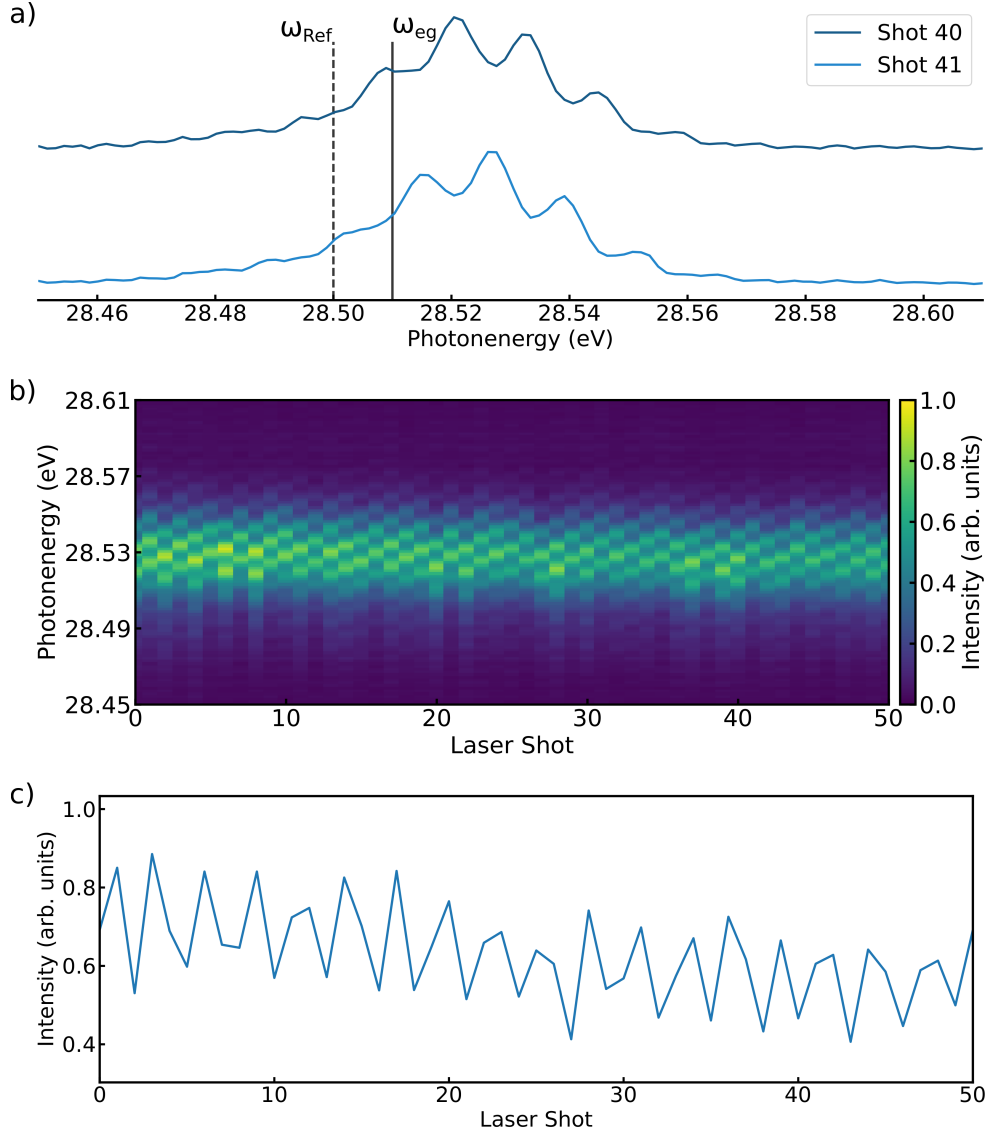}
\captionsetup{format=hang}
\caption{Retrieval of the XUV reference signal. 
(a) Two-pulse interference spectrum for two consecutive laser shots at a fixed pump-probe delay $\tau=300$\,fs. 
Solid line: optical resonance probed in the experiment ($\omega_{eg}$), dashed line: position at which the signal phase is evaluated ($\omega_\mathrm{ref}$). 
(b) Shot-to-shot phase modulation of 50 consecutive laser shots at a pulse delay of $\tau=300$\,fs and FEL repetition rate of 50\,Hz.
(c) Modulated signal extracted at 28.50\,eV for a modulation of 18.24\,Hz, sampled at a rate of 50\,Hz.  
}
\label{fig2}
\end{figure}
Secondly, extraction of the reference signal from the XUV beamline spectrometer (2D CCD array detector) becomes possible, which allows tracking the XUV phase fluctuations $\delta \alpha$ on the single-shot level (Fig.\,\ref{fig1}d). 
Fig.\,\ref{fig2}a shows the XUV beamline spectrometer signal for two consecutive shots, revealing a clear interference fringe spectrum of the XUV pump-probe pulses. 
In the Ramsey-type interference fringes, the fringe spacing is inversely proportional to the pump-probe delay $\tau$, while the phase-offset corresponds to the CEP difference between both pulses. 
Shot-to-shot modulation of the relative phase between the seed laser pulses leads to a modulation of the XUV interferogram at $n$-times the modulation frequency (Fig.\,\ref{fig2}b). 
Hence, while keeping $\tau$ fixed, extracting the amplitude of the spectra at one specific photon energy $\hbar \omega_\mathrm{ref}$ for multiple consecutive laser shots, returns a periodic oscillating waveform suitable for the lock-in demodulation (Fig.\,\ref{fig2}c). 
To minimize the influence of FEL intensity fluctuations on the extracted phase, we normalized the area of each interferogram before extracting the phase.

For a direct comparison of the signal demodulation using the commercial LIA with Ref1 and the ULIA with Ref2, we apply both methods to the same data set, that is the $\mathrm{3s^{2}3p^{6} \rightarrow 3s^{1}3p^{6}6p^{1}}$ Fano resonance in argon atoms, as probed in Ref.\,\cite{wituschek_tracking_2020}. 
In the experiment, the FEL was operated on the 6th harmonic of the seed laser (28.53\,eV) and the XUV pump-probe delay was scanned from 150\,fs to 600\,fs in 2\,fs steps. 
The phase modulation frequency was $\Omega = 18.24$\,Hz (at the 6th harmonic) and at each delay the signal was demodulated over 800 laser shots when using the commercial LIA and Ref1, and 700 laser shots when using the ULIA and Ref2, both at a repetition rate of 50\,Hz. 
In the previous method, the XUV interference data was extracted by gating the ion-TOF spectra on the Ar$^+$ mass with a boxcar integrator yielding a sufficiently smooth signal for demodulation with the commercial LIA. 
The reference was extracted from the optical interference in the seed laser interferometer (Ref1 in Fig.\,\ref{fig1}d). 
In the current work, we use the same data set, but demodulate the data with the ULIA algorithm and use the XUV beamline spectrometer data for the reference (Ref2 in Fig.\,\ref{fig1}d). 

\section{Results}
\label{sec:results}
\begin{figure}[htb!]
\centering
\includegraphics[width=0.8\linewidth]{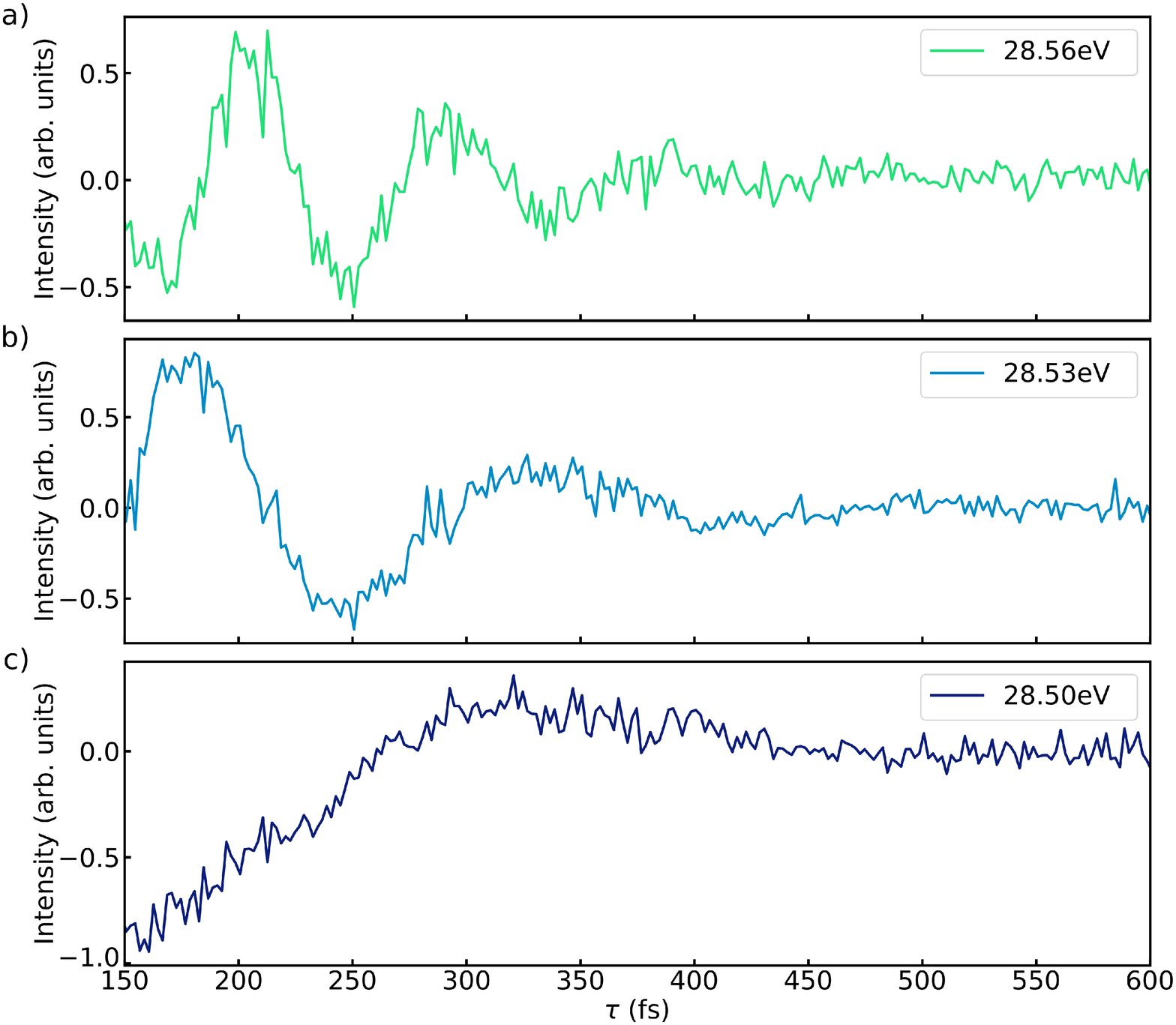}
\captionsetup{format=hang}
\caption{Demodulated time-domain interferograms using different references extracted at a photon energy of 28.56\,eV (a), 28.53\,eV (b), and 28.50\,eV (c) from the XUV spectrometer. 
The fringe frequency reduces clearly for reference signals extracted at photon energies closer to the transition frequency of the Fano resonance frequency (28.51\,eV).
}
\label{fig3}
\end{figure}

First, we demonstrate the principle of rotating frame detection. 
The excited Fano resonance is at 28.51\,eV, which corresponds to an interference fringe spacing of 145\,as in the time domain signal and would require an extremely fine sampling with $\tau$-increments of $< 145/2$\,as.  
In the PM technique, beating the signal with the reference leads to a down-shifted frequency of $\Delta \omega$, which depends on the reference frequency $\omega_\mathrm{ref}$. 
In the current work, $\omega_\mathrm{ref}$ can be freely chosen within the bandwidth of the XUV pulse spectrum (Fig.\,\ref{fig2}a), enabling the optimization of the signal frequencies to the experimental conditions. 
Figs.\,\ref{fig3}a--c exemplify the demodulated signal for references Ref2 extracted at three different frequencies $\omega_\mathrm{ref}$, clearly demonstrating the flexibility in arbitrarily down-shifting the signal frequency. 
The corresponding interference fringe periods are 106\,fs, 143\,fs, and 459\,fs, which correspond to a frequency down-shift of up to a factor of 3167, significantly simplifying data acquisition.  
This flexibility was not available in the previous detection method\,\cite{wituschek_tracking_2020}, where $\omega_\mathrm{ref}$ was defined by a continuous-wave (cw) metrology laser tracing the seed laser interferometer and was therefore not tunable. 
The cw laser was necessary to obtain a smooth waveform for the lock-in demodulation with the commercial LIA\,\cite{bruder_phase-synchronous_2018}. 
Here, a frequency down-shift by a factor of 51 was achieved, almost two orders of magnitude less than in the current work. 

\begin{figure}[htb!]
\centering
\includegraphics[width=0.75\linewidth]{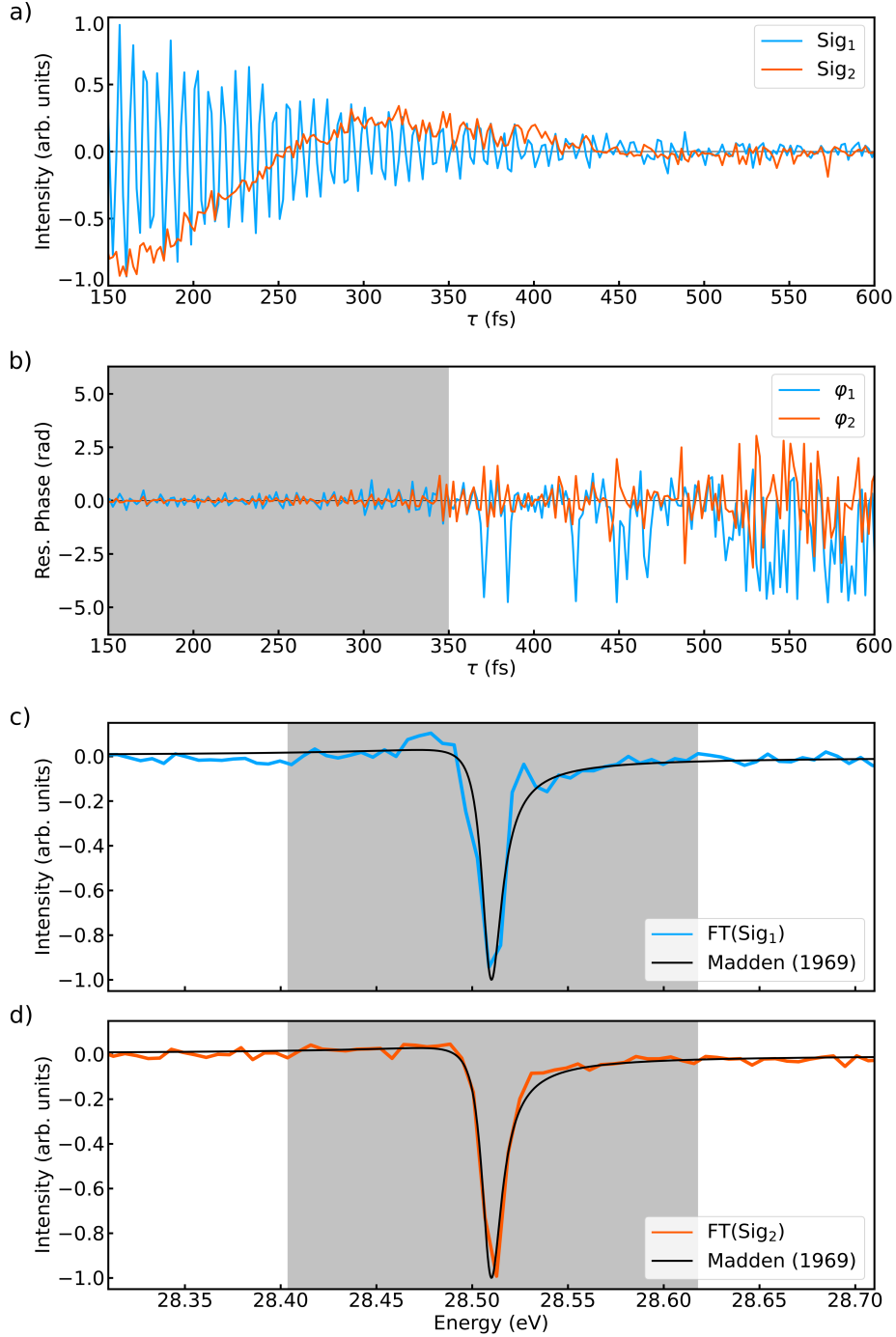}
\captionsetup{format=hang}
\caption{Comparison between the phase stabilization methods using Ref1 (blue) and Ref2 (orange) (cf. Fig.\,\ref{fig1}d) for demodulation. 
(a) Time-domain interferograms of the Fano resonance. Both signals exhibit strongly down-shifted frequencies. Due to the different reference signals, the oscillation periods differ and are 7.5\,fs (blue trace) and 491\,fs (orange trace), respectively. 
(b) Residual phase fluctuations in the demodulated time-domain signals. 
(c, d) Fourier spectra (real part) of the time domain signals along with a steady-state absorption spectrum obtained with synchrotron radiation~\cite{madden_resonances_1969} (black).
}
\label{fig4}
\end{figure}

This clear difference between the two demodulation methods using the fixed Ref1 and the flexible Ref2 is shown in Fig.\,\ref{fig4}. 
Fig.\,\ref{fig4}a compares the corresponding down-shifted signal frequencies in the time domain. 
For a quantitative comparison of the signal quality, the phase noise is shown in Fig.\,\ref{fig4}b. 
For a better visualization, the theoretically expected linear phase slope is subtracted from the data revealing the phase fluctuations relative to an ideal harmonic oscillation. 
At a delay of $> 350$\,fs where the signal amplitude has decayed to $\approx 18$\% the residual phase jitter clearly increases. 
In this region, intensity fluctuations of the FEL start to dominate the signal and a comparison of the residual phase noise from both demodulation methods is not reasonable. 
In the region $< 350$\,fs (gray shaded), the phase noise is low for both methods, meaning that the phase for these delays is stable.
However, the RMS fluctuations in this region are a factor of 1.4 smaller for the ULIA demodulation. 
This suggests, that using the XUV reference (Ref2, cf. Fig.\,\ref{fig1}b) for the signal demodulation, indeed improves the signal quality. 

Next, we evaluate the Fourier spectra of both signals. 
The interfering ionization pathways at a Fano resonance alter the absorption line shape between absorptive and dispersive character depending on the phase shift between the direct and indirect ionization path\,\cite{fano_effects_1961}. 
The Fano line shape is recovered in the real part of the Fourier transforms of the WPI signals. 
Accordingly, Fig.\,\ref{fig4}c,d compare the Fourier spectra of both signals, up-shifted by the corresponding $\omega_\mathrm{ref}$-values to obtain the absolute frequency axis. 
To recover the line shape in the lock-in detection correctly, a calibration of the absolute phase shift between signal and reference has to be performed, which was not done for the demodulation using the ULIA. 
Instead, in Fig.\,\ref{fig4}d the phase parameter was optimized by fitting the data to the reference spectrum taken from a static absorption spectrum measured with synchrotron radiation~\cite{madden_resonances_1969}. 
For a better comparison between both demodulation methods, the same fit procedure was applied to the data in Fig.\,\ref{fig4}c. 

\begin{figure}[ht!]
\centering
\includegraphics[width=0.75\linewidth]{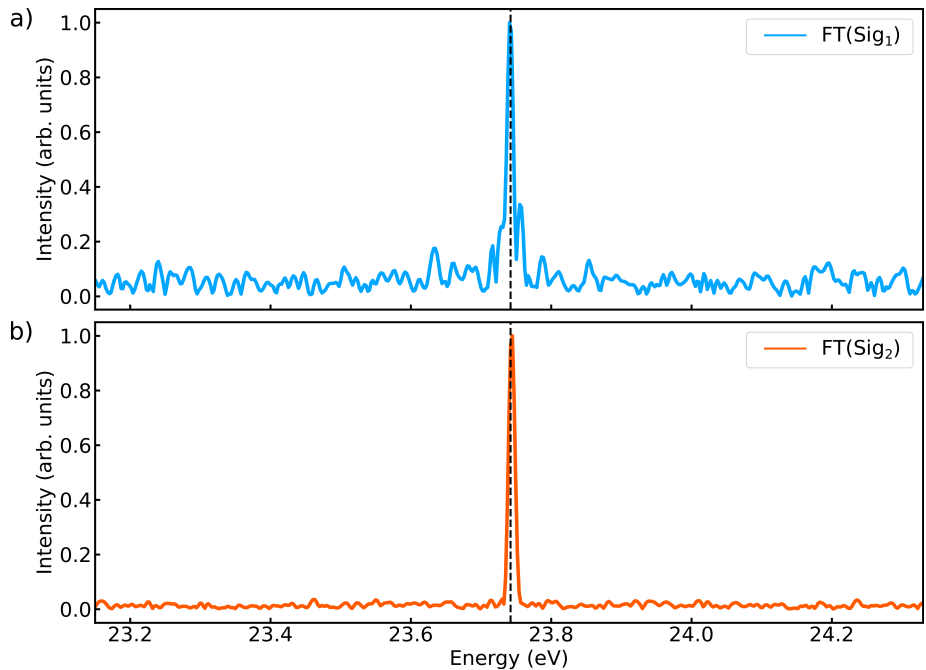}
\captionsetup{format=hang}
\caption{Fourier spectra of the $1s^2 \rightarrow 1s4p$ coherence excited in He atoms for demodulation using Ref1 (a) and Ref2 (b). 
The overall SNR in (b) has been improved by a factor of 3.1 compared to (a). 
Moreover, in (a) the spectral peak exhibits sidelobes as a typical sign for discontinuities in the temporal phase of the signal. 
These effects are well compensated when using Ref2 for the demodulation.
}
\label{fig5}
\end{figure}

Comparing both demodulation methods with the synchrotron data, a better qualitative match is found for the phase stabilization using the XUV reference Ref2 (Fig.\,\ref{fig4}d). 
For a quantitative comparison, we calculated the RMS deviation from the synchrotron data in the gray shaded spectral region. 
Outside of this region, no significant signal is expected and the spectra are dominated by statistical noise. 
The RMS deviations are $\mathrm{4.93\cdot10^{-3}}$ (Fig.\,\ref{fig4}c) and $\mathrm{0.96\cdot10^{-3}}$ (Fig.\,\ref{fig4}d), respectively, implying a signal improvement by a factor of 5.1 when using Ref2 for the signal demodulation. 
This is in accordance with the time domain evaluation, showing the same tendency of an improved signal quality when using the new stabilization scheme. 
The result is also in agreement with another experiment (Fig.\,\ref{fig5}), where we measured the coherence between two bound states in helium ($1s^2 \rightarrow 1s4p$, 23.74\,eV)\,\cite{wituschek_tracking_2020}. 
Here, we compare the signal-to-noise ratio (SNR) obtained with both stabilization methods, yielding an SNR of 14.0 using Ref1 and 43.3 using Ref2 for demodulation, which is a remarkable high SNR for interference experiments in the XUV domain. 
For the SNR determination we defined the noise floor as the mean value of the background signal plus two times its standard deviation. 

In general, using the XUV reference (Ref2) for demodulation one may expect an even higher improvement of the signal quality than observed in our study. 
The limited performance can be explained by two reasons. 
First, the shot-to-shot amplitude fluctuations of the FEL output introduces noise which is not corrected by either of the two stabilization methods. 
Second, the correlation of the phase fluctuations $\delta \alpha$ is small between consecutive laser shots. 
Hence, the bandwidth of the lock-in-based stabilization method is insufficient to correct these high-frequency phase fluctuations. 
The correction of such high-frequency noise contributions requires single-shot self-referencing methods, e.g. as implemented in Refs.\,\cite{maroju_attosecond_2020, maroju_analysis_2021}. 
We investigated the possibility of single-shot self-referenced data correction using the XUV spectrometer data and adapting the ULIA algorithm accordingly. 
However, this did not improve the data quality further due to the large uncertainties for extracting the required amplitude, phase and delay values from the XUV spectrograms. 

\section{Conclusion}
\label{sec:conclusion}
We introduced a new stabilization scheme for interferometric XUV pump-probe experiments with seeded FELs. 
The method is based on a previous stabilization scheme combining phase modulation and lock-in amplification to stabilize the seed laser interferometer at the fundamental wavelength\,\cite{wituschek_tracking_2020}. 
In the current work, we have extended the method to include the stabilization of the XUV beampath, in particular of phase fluctuations introduced in the HGHG process. 
Applying both methods to track the coherence decay of a Fano resonance in argon atoms, we find a clear signal improvement with the new scheme. 
Further improvement of the signal quality is limited by high-frequency noise introduced in the HGHG process which cannot be suppressed with the limited bandwidth of the stabilization method, and by the amplitude fluctuations which are not corrected in the stabilization scheme. 

\section{Acknowledgements}
We gratefully acknowledge the support of the FERMI staff, in particular of the machine team and the laser team. 

\section{Funding}
Bundesministerium für Bildung und Forschung (BMBF, 05K16VFB);
European Research Council (ERC) with the Advanced Grant “COCONIS” (694965);
Deutsche Forschungsgemeinschaft (DFG) RTG2079 and STI125/19-2; 
Swedisch Research Council (2018-03731); Knut and Alice Wallenberg Foundation (2017.0104).

\section{Disclosures}
The authors declare no conflicts of interest.

\section{Data Availability Statement}
Data is available from the corresponding author upon reasonable request.
\\

\bibliography{FERMI_ULIA}

\end{document}